\documentstyle[12pt,aasms4,psfig]{article}
\def\gte{\,\lower.6ex\hbox{$\buildrel >\over \sim$} \, }
\def\lte{\,\lower.6ex\hbox{$\buildrel <\over \sim$} \, }

\received{}
\accepted{}
\journalid{}{}
\articleid{}{}

\lefthead{M.P. D\c{a}browski and M.A. Hendry}
\righthead{Non-Uniform Pressure Universes}

\begin{document}

\title{The Hubble Diagram of Type Ia
Supernovae in Non-Uniform Pressure Universes}

\author{Mariusz P.~D\c{a}browski$^{1,2}$} 

\affil{$^1$Astronomy Centre, University of Sussex, Falmer, Brighton BN1 9QH, 
UK\\$^2$Institute of Physics, University of Szczecin, Wielkopolska 15, 
70-451 Szczecin, Poland}

\author{Martin A.~Hendry,$^{2,3}$}
\affil{$^1$Astronomy Centre, University of Sussex, Falmer, 
Brighton BN1 9QH, UK\\
$^3$Dept of Physics and Astronomy, University of Glasgow, Glasgow G12 8QQ, UK}

\begin{abstract}
We use the redshift-magnitude relation, 
as derived by D\c{a}browski (1995), for the two exact
non-uniform pressure spherically symmetric Stephani universes with the observer positioned at the center of symmetry, to test the
agreement of these models with recent observations of high redshift type Ia
supernovae (SNIa), as reported in Perlmutter et al. (1997). By a particular 
choice of model parameters, we show that these models can give an excellent fit
to the observed redshifts and (corrected) B band apparent magnitudes of the
Perlmutter et al. data, but for an age of the Universe which is
typically about two Gyr -- and may be more than three Gyr -- greater than in
the corresponding Friedmann model, for which non-negative
values of the deceleration parameter appear to be favoured by the data.
We show that this age increase is obtained for a wide range of the non-uniform 
pressure parameters of the Stephani models. We claim this paper is the first 
attempt to compare inhomogeneous models of the universe with real 
astronomical data.

Several recent calibrations of the Hubble parameter, from the Hubble diagram
of SNIa and other distance indicators, indicate a value of $H_0 \simeq 65$,
and a Hubble time of $\sim 15$ Gyr. Based on this value for $H_0$ and
assuming $\Lambda \geq 0$, the P97 data would imply a Friedmann age of at
most 13 Gyr and in fact a best-fit (for $q_0 = 0.5$) age of only 10 Gyr.
Our Stephani models, on the other hand, can give a good fit to the P97 data
with an age of up to 15 Gyr. The Stephani models considered here
could, therefore, significantly alleviate the conflict between recent
cosmological and astrophysical age predictions. The choice of model
parameters is quite robust: in order to obtain a good
fit to the current P97 data, one requires only that the non-uniform pressure 
parameter, $a$, in one of the models is negative and satisfies 
$|a| \lte$  3 km$^2$ s$^{-2}$ Mpc$^{-1}$. This limit gives a value for the 
acceleration scalar, $\dot{u}$, 
of order $\vert \dot{u} \vert \lte 0.66 \times 10^{-10} \, r$ Mpc$^{-1}$, 
where $r$ is the radial coordinate in the model. Thus, although the pressure is 
not zero at the center of symmetry, $r = 0$, the effect of acceleration is 
non-detectable at the center since the acceleration scalar vanishes there. 
However, the effect of 
the non-uniform pressure on the redshift-magnitude relation is clearly seen 
since neighbouring galaxies are not situated at the center and they 
necessarily experience acceleration. By allowing slightly
larger, negative, values of $a$ one may `fine tune' the model to give an
even better fit to the P97 data.
\end{abstract}


\section{Introduction}

The standard isotropic Friedmann cosmological models have naturally been the
most widely investigated models in studies of the large-scale structure of
the Universe. This is hardly surprising, in view of their mathematical
simplicity and their generic prediction of an approximately linear Hubble
expansion at low redshift, which is in excellent agreement with observational
data (c.f. Strauss \& Willick 1995; Postman 1997).
Even in Friedmann models, however, the relation between apparent magnitude
and log redshift is in general non-linear at higher redshift and depends
explicitly on the spatial curvature of the Universe -- or equivalently on the
deceleration parameter, $q_0$.

For several decades astronomers have attempted to use the Hubble diagram of
some suitable `standard candle' (e.g. first-ranked cluster galaxies) to place
constraints on the global geometry of the universe by comparing the observed
redshift-magnitude relation of the standard candle with that predicted in
Friedmann models with different values of $q_0$
(c.f. Peach 1970; Gunn \& Oke 1975; Schneider, Gunn \& Hoessel 1983;
Sandage 1988). Results from such
analyses have thus far proved inconclusive, however. Due to the intrinsic
dispersion in the luminosity function of the standard candles available, 
one previously had to reach at least $z \simeq 1$ before the predictions of 
models with different
values of $q_0$ became sufficiently distinct to be detectable; at the same
time, however, the effects of luminosity and number density evolution also
become important at these redshifts, and are very difficult to correct for.
The situation for the Hubble diagram of quasars is equally -- if not even
more -- problematical. Tonry (1993) suggested that the contraints on $q_0$
from such studies were no better than $-1 < q_0 < 1$.

Recently, however, it has been suggested that type Ia supernovae (hereafter
SNIa) represent a standard candle of sufficiently small dispersion to allow
meaningful estimates of $q_0$ now to be derived from the SNIa Hubble diagram 
at more moderate redshift. In Perlmutter et al. (1997; hereafter P97) a 
preliminary analysis is presented
of seven distant SNIa in the range $0.35 < z < 0.50$. A comparison of the
SNIa magnitudes and redshifts with the predicted relation for various
Friedmann models appears to exclude large
negative values of $q_0$, and is best fitted
by values close to $q_0 = 0.5$. This poses a potentially serious problem for
Friedmann models. Since many recent determinations of the Hubble constant
(including a number of analyses using SNIa) suggest that $H_0$ lies in the
range 65 -- 70, this would imply an age of the universe of less than 10 Gyr
in the `standard' $\Omega_0 = 1, \Lambda = 0$ scenario. This result would
appear to be in sharp conflict with recent astrophysical age 
determinations from e.g. globular clusters and white dwarf cooling
(c.f. Chaboyer 1995; Hendry \& Tayler 1996) -- a conflict which is only
slightly alleviated by revisions to globular cluster age estimates in the light
of results from the HIPPARCOS satellite (Chaboyer et al. 1997).
Reducing the value of $\Omega_0$ lessens the conflict somewhat, but
agreement is still only marginal if one accepts a robust lower bound for the
matter density of $\Omega_{\rm{m}} = 0.3$, as has been suggested by
several different methods of analysing large-scale galaxy redshift surveys
(c.f. Strauss \& Willick 1995). This situation has helped to give a
renewed impetus to models with a positive cosmological 
constant (c.f. Liddle et al. 1996) which contributes an
additional component, $\Omega_\Lambda$, to make up the critical density and
at the same time extends the age of the Universe by up to 2 Gyr -- depending
on the value of $H_0$ and $\Omega_{\rm{m}}$. However, because of the relation
\begin{equation}
q_0 = \frac{1}{2} \Omega_{\rm{m}} - \Omega_\Lambda  ,
\end{equation}
it is clear that a positive value of $q_0$ is incompatible with a
positive value of $\Lambda$ unless the matter density is at
least two-thirds of the critical density. The $q_0 = 0$ case, assuming
$\Omega_{\rm{m}} = 2/3$, $\Omega_{\Lambda} = 1/3$ and
$H_0 = 65$, would give an age of the Universe of just over 11 Gyr;
as $q_0$ increases the age is decreased still further. Thus, if the 
results of P97 prove to be correct and the deceleration
parameter {\em is\/} non-negative, then the conflict between cosmological and
astrophysical age predictions remains firmly unresolved -- at least if
$H_0 \gte 65$. Independent results showing that a positive value of $\Lambda$ 
is incompatible with the so-called VLBI data (Kellermann 1993), using the
angular diameter test, were obtained by Krauss \& Schramm (1993) and 
Stelmach (1994). 

In this paper we propose one method to alleviate this age conflict by
considering some inhomogeneous cosmological models in which the relation
between the age of the Universe and a generalised Hubble constant is more
general than in the Friedmann case. Despite some theoretical plots of the 
observational quantities for inhomogeneous models 
(e.g. Goicoechea \& Martin-Mirones 1987, Moffat \& Tatarski 1995, 
D\c{a}browski 1995, Humphreys, Maartens \& Matravers 1997) this paper is --
as far as we are aware -- the first
to compare these models with real astronomical data. In particular, 
we show that taking an inhomogeneous model into account allows us to obtain 
a good fit between the predicted redshift-magnitude
relation and P97 data, but for an age of the Universe which
is several Gyr older than in the Friedmann case. The models under
consideration have been discussed before and are known as Stephani
Universes (c.f. Kramer et al. 1980; Krasi\'nski 1983; 
D\c{a}browski 1993). In these models the energy density $\varrho$ depends 
just on the cosmic time, similarly to the Friedmann models,
but the pressure, $p$, is a function of both spatial coordinates and a time 
coordinate; hence the
models are usually referred to as `inhomogeneous pressure universes'. In the 
spherically symmetric case under consideration, the pressure is just a function 
of time and radial coordinate which means that its values are the same on 
spheres, $r =$ constant, around the center of symmmetry but differ from 
sphere to sphere. This essentially means that there is a spatial pressure 
gradient and particles are accelerated in the direction from high-pressure 
regions to low-pressure regions. 
This effect is usually described by the acceleration vector $\dot{u}_r$ which 
in the case of spherical symmetry has only one (radial) component, or by the 
acceleration scalar $\dot{u}$ (cf. Eqs. 2.18 and 2.23 of D\c{a}browski 1995). 
The acceleration represents the 
combined effect of gravitational and inertial forces on the fluid which, 
in fact, similarly as in Newtonian physics, are unable to be separated. As a first approximation we assume the observer is placed at the center
of symmetry, which results in no pressure gradient at the observer's postion.
This, in a sense, contradicts the Copernican Principle, but can easily be
overcome by applying the formulae for a non-centrally observer given in
Section 5 of D\c{a}browski (1995) -- an appropriate generalisation once larger
samples of high-redshift supernovae become available.

In the standard approach we neglect the effect of pressure (i.e. we take 
pressureless dust -- with pressure $p = 0$) as we evaluate chaotic velocities 
of galaxies to be small. This results in taking the acceleration, $\dot{u}$, 
also to be zero (cf. Ellis, 1971, for a discussion of the relation between 
these quantities). However, if there was large flux of neutrinos or 
gravitational waves for instance, this assumption
would not be correct and we would need 
to take radiation pressure ($p = \frac{1}{3} \varrho$) into account. 
This has been of 
course investigated for isotropic cosmologies (D\c{a}browski \& Stelmach 1986, 
1987) and all the observational quantities have been found.  Our main point 
here is, however, that early universe processes such as phase 
transitions (for details see Vilenkin 1985, Kolb \& Turner 1990, 
Vilenkin \& Shellard 1994) may result in having different exotic types of 
matter (e.g. cosmic strings) with many different types of 
equations of state. In the easiest case (straight cosmic strings) they may
end up with the exotic equation of state 
$p = - \frac{1}{3} \varrho$, but in general, 
the equation of state can be more complicated (e.g. Vilenkin \& Shellard 1994 
and de Vega \& Sanchez 1994, in the context of superstrings) or 
spatially dependent (e.g. Narlikar, Pecker \& Vigier 1991). The latter 
would especially be 
the case of our interest. Of course the standard energy conditions of 
Hawking and Penrose might be violated (cf. Hawking \& Ellis 1973) which also
happens for inflationary models for instance and our considerations here 
are, in a sense, on the same footing as 
those phenomena. As for the Stephani models, which do not admit any 
global barotropic equation of state, it has been shown that there exists 
a consistent nonbarotropic equation of state 
and the full thermodynamical scheme exists 
(Quevedo \& Sussman 1995, Krasi\'nski, Quevedo \& Sussman 1997).

Regardless of the physical background of the models under consideration, 
one of our main tasks here is to draw attention to the entire class of 
inhomogeneous models which could be a 
useful alternative to Friedmann models in helping to resolve the apparent
incompatibility of measurements of the Friedmann cosmological parameters. 
Even if the final outcome (after a thorough comparison with data) shows that
the 
universe is indeed isotropic and homogeneous this conclusion must be 
drawn by applying some `averaging scale-dependent procedures' 
( cf. Ellis 1984, Buchert 1997) since we evidently cannot see the universe 
to be like that on smaller scales. Being spherically symmetric, 
Stephani models can also be applied to a local underdense/overdense spherical 
region embedded in a globally isotropic Friedmann universe 
(Moffat \& Tatarski 1995) in some analogy to the 
so-called `Swiss Cheese' model (Kantowski, Vaughan \& Branch 1995).

The reader interested in more generic models should be referred to the recent
review by Krasi\'nski (1997), as well as to some earlier papers
concerning the most popular generalization of the Friedmann models
such as the spherically symmetric Tolman Universes which are inhomogeneous 
density pressure-free dust shells (Tolman 1934; Bondi 1947; Bonnor 1974). 
Their properties have been 
studied quite thoroughly in Hellaby \& Lake (1984, 1985) and Hellaby 
(1987, 1988) and the observational relations for Tolman models
were studied by Goicoechea \& Martin-Mirones (1987), Moffat \& Tatarski
(1995) and quite recently by Humphreys, Maartens \& Matravers (1997). 
However, in none of these cases has a comparison with real astronomical data
been carried out.

The outline of this paper is as follows.
In Section 2 we reproduce the redshift-magnitude relations for the two
Stephani models considered here, as recently derived in D\c{a}browski
(1995). In Section 3 we briefly describe the SNIa data of P97.
In Section 4 we fit these data to the redshift-magnitude relations of both
Friedmann and Stephani models and thus obtain best-fit values for the
model parameters. We then discuss the results of these fits and compare the
age of the Universe given by the best-fit model parameters in the
Friedmann and Stephani cases. Finally in Section 5 we summarise our
conclusions.

\section{The Redshift-magnitude relation for inhomogeneous pressure models}
Recently D\c{a}browski (1995) has considered the redshift-magnitude relation
for Stephani universes. Two exact cases were presented and the predicted
relations were plotted for a range of different parameter values. The
relations were defined following the method of Kristian \& Sachs (1966),
of expanding all relativistic quantities in power series and truncating at
a suitable order. Approximate formulae, to first order in redshift $z$, for 
Model I and Model II respectively were given by
\begin{eqnarray}
m_{\rm{B}} & = & M_{\rm{B}} + 25 
+ 5\log_{10}{ \left[cz \left(\frac{a\tau_{0}^2 + b\tau_{0} + d}
{2a\tau_{0} + b}\right)\right]} \nonumber \\
& + & 1.086 \left[ 1 + 4a \frac{\left( a\tau_{0}^2 + b\tau_{0} + d 
\right)}{\left(2a\tau_{0} + b \right)^2} \right] z
\end{eqnarray}
and 
\begin{eqnarray}
m_{\rm{B}} & = &  M_{\rm{B}} + 25 - 5\log_{10} {\frac{2}{3\tau_{0}}} + 
5\log_{10}{cz}
\nonumber \\
      & + & 1.086 \left( \frac{1}{2} + \frac{9}{8} c^2 \alpha
\tau_{0}^{\frac{4}{3}} \right) z   ,
\end{eqnarray}
which are essentially equations (5.6) and (5.10) respectively of
D\c{a}browski (1995).
\footnote{But written now for B band observations, which are the
case under study in this paper. Note also that we have corrected a
typographical error which appeared in D\c{a}browski (1995) -- the factor of 4 
in the final term of Eq. (2) replacing the factor of 2 in Eq. (5.6) of that
paper. We thank Chris Clarkson for drawing our attention to this error.}
Here $m_{\rm{B}}$ and 
$M_{\rm{B}}$ denote apparent and absolute magnitude respectively
in the B band, $\tau_{0}$ denotes the current age of the Universe and
$c = 3 \times 10^5$ km s$^{-1}$ is the velocity of light. The constants
$a, b$ and $d$ are parameters of Model I and $\alpha$ is a parameter
of Model II. Convenient units for these parameters are:
$[a]$ = km$^2 \,$s$^{-2}\,$ Mpc$^{-1}$,
$[b]$ = km s$^{-1}$, $[d]$ = Mpc
and $[\alpha c^2]$ = (km s$^{-1}\,$ Mpc)$^{-\frac{4}{3}}$. 
From the definition of the acceleration scalar 
(c.f. D\c{a}browski 1995)
we conclude that the parameters which relate directly 
to the non-uniformity of the pressure are $a$ in Model I and
$\alpha$ in Model II. In Model I, $b$ plays a similar role to the
coefficient of time in the expression for the scale factor 
($R \propto \tau^p$ where $p$ is any power) in Friedmann models and can be 
considered as an immanently Friedmannian parameter of the Stephani models, 
while $a$ and $\alpha$ are completely non-Friedmannian. 

The models considered here are spherically symmetric, which means that we
can have both centrally placed and non-centrally placed observers. For
simplicity the 
redshift-magnitude relations reproduced above correspond to a centrally
placed observer.  D\c{a}browski (1995) also derived relations for the case
of a non-centrally placed observer which, although more general,
introduced several additional free parameters. The main difference in this
more general case is that the apparent magnitude depends on the
position of the source in the sky, and renders comparison
with the Friedmann case more complicated. We thus consider only
centrally placed observers in this paper.

Note that these formulae are truncated at first order in $z$ and thus
would become increasingly inaccurate if applied to redshifts greater
than or equal to unity. Since the redshifts of the P97
preliminary data extend only to $z \sim 0.5$ we proceed with
the first-order expressions, and will also use the first-order
expression for Friedmann models when comparing the fits. In future
work, as the database of high redshift SNIa grows, we will extend
the approximate redshift-magnitude relations to higher order, as required.

The reader is referred to  D\c{a}browski (1995) for a detailed discussion
of the derivation of the above formulae. Note, however, that for Model I
the expression for the generalised scale factor, $R(\tau)$, as a function
of cosmic time, $\tau$, is given by (Eq. 2.11 of  D\c{a}browski 1995)
\begin{equation}
R(\tau) = a \tau^2 + b \tau + d   .
\end{equation}
However, $R(\tau)$ does not have to be positive (D\c{a}browski 1993, 
Section IV.A and Fig. 6) for the Stephani models. For the subclass under 
consideration, $R(\tau)$ easily relates to the spatial curvature of the models 
\begin{equation}
k(\tau) = - 4 \frac{a}{c^2} R(\tau)   ,
\end{equation}
and the curvature index is not constant in time as for Friedmann models. 
In principle, one can restrict $R(\tau)$ to be positive (this is especially 
reasonable, if we want to 
obtain the full Friedmann limit), which ends up with the simple relation 
for the spatial curvature of the models being positive for negative non-uniform 
pressure parameter $a$ and negative for positive $a$. Since $R(\tau) = 0$ at 
the singularity (the Big-Bang) we can require that $R(\tau) \rightarrow 0$ as 
$\tau \rightarrow 0$
(i.e. we set the origin of our time coordinate at the Big Bang) and thus
demand that $d$ is identically zero. Therefore, according to condition (2.13) 
of D\c{a}browski (1995), which, in fact, allows
one to have the Friedmann limit for the Stephani models under consideration, 
we have 
\begin{equation}
b^2 = 1   ,
\end{equation}
for the values of $b$. Without loss of generality we assume that $b = +1$ 
(cf. the discussion above about the meaning of $b$ in Friedmann models) -- 
leaving only one free parameter of the Model I, which is $a$. 

Other important physical quantities of Model I are as follows: the energy density 
\begin{equation}
\frac{8\pi G}{c^4} \varrho(\tau) = \frac{3}{(a\tau^2 + b\tau)^2}   ,
\end{equation}
the pressure 
\begin{equation}
\frac{8\pi G}{c^2} p(\tau) =  - \frac{1}{(a\tau^2 + b\tau)^2} \left[ 1 + 2a 
\left( a\tau^2 + b\tau \right) r^2 \right]   ,
\end{equation}
and the acceleration scalar 
\begin{equation}
\dot{u} = - 2 \frac{a}{c^2} r   .
\end{equation}
From the above one can see that the finite density singularities of pressure 
appear at $ r \to \infty$ \footnote{Note that in Stephani models we use the 
so-called isotropic radial coordinate which results in 
the Friedmann metric to be taken in the 
isotropic form. It relates to the commonly used nonisotropic coordinate $r_{N}$ 
via the relation $r_{N} = r/(1 + (1/4)r^2)$ (for details see Section II of 
D\c{a}browski 1993 ). Because of isotropy at every point one cannot 
differentiate the centers $r = 0$ and $ r = \infty$ in a Friedmann universe, 
which is not the case for the 
Stephani models.} where there is the antipodal center of symmetry. We assume
that we are placed at the center of symmetry at $r = 0$ so we have 
these singularities far away from us (see Fig. 6 of D\c{a}browski 1993). 
Of course we cannot live at the singularity of pressure. 

At the center of symmetry the fluid fulfils the barotropic equation of 
state $p = - \frac{1}{3} \varrho$ (as for straight cosmic strings, cf. Vilenkin 
1985), while at $ r \to \infty$ the pressure goes to either plus or minus 
infinity. Then, 
assuming $R(\tau) > 0$, it diverges to minus infinity if $a > 0$, and to plus 
infinity if $a < 0$. In such a case, the particles are accelerated away from a 
high pressure region at $r = 0$ to low pressure regions at $r  \neq 0$, 
if $a > 0$, and toward a low pressure region at $r = 0$ from high pressure 
regions at $r \neq 0$, if $a < 0$. Of course if $R(\tau) < 0$ the situation is 
the opposite. The acceleration scalar is zero at $r = 0$ and it diverges at 
$ r \to \infty$.

In the case of Model II the time-dependent curvature index is given by 
($\beta$ plays here the same `Friedmannian' role as $b$ in Model II)
\begin{equation}
k(\tau) = - \alpha\beta^2 c^{-2} \tau ^{\frac{2}{3}}   ,
\end{equation}
while the energy density, pressure and acceleration scalar are
given respectively by (D\c{a}browski 1993, Appendix C) 
\begin{equation}
\frac{8 \pi G}{c^4} \varrho(\tau) = \frac{4}{3} \frac{1}{\tau^2} - 
\frac{3 \alpha}{\tau^{\frac{2}{3}}}  ,
\end{equation}
\begin{equation}
\frac{8 \pi G}{c^2} p(\tau) = \frac{2 \alpha}{\tau^{\frac{2}{3}}} - \frac{4}{3} 
\frac{\alpha \beta^2}{\tau^{\frac{4}{3}}} r^2 + \alpha^2 \beta^2 r^2   ,
\end{equation}
and
\begin{equation}
\dot{u} = - \frac{1}{2} \alpha \beta r   .
\end{equation}

\section{The SNIa observations of Perlmutter et al. (1997)}

Type Ia supernovae are thought to be the result of the thermonuclear disruption
of a white dwarf star which has accreted sufficient matter from a binary
companion to reach the Chandrasekhar mass limit. For several decades they
have been considered as suitable (nearly) standard candles for the testing
of cosmological models because of the relatively small dispersion of their
luminosity function at maximum light and the fact that they are observable
at very great distances. In recent years the Hubble diagram of SNIa has been
used by a number of authors to obtain estimates of the Hubble constant
(c.f. Riess, Press \& Kirshner 1996; Hamuy et al. 1995, 1996; Branch et al. 
1996) and the motion of the Local Group (Riess, Press \& Kirshner 1995).
P97 consider the redshift-magnitude relation of SNIa at
high redshift, observed by the `Supernova cosmology project',
as a means of constraining $q_0$. In P97 SNIa are not treated as precise
standard candles, but a `stretch factor' correction is applied to account 
for the correlation between SNIa luminosity and the shape of their light curve.

In this paper we use the redshifts and $B$ band magnitudes -- with and without
light curve shape corrections -- as presented in Table 1 of P97, to which
the reader is referred for details of their observing strategy, data
reduction procedures and magnitude error estimates.

\section{Comparison of the data with Friedmann and Stephani Models: results
and discussion}

\subsection{Friedmann Models}

Figure 4 of P97 shows the Hubble diagram of their SNIa compared with
the theoretical magnitude-redshift relations for a Friedmann model with 
different combinations of $\Omega_{\rm{m}}$ and $\Omega_\Lambda$. While P97
argue correctly that one should generally express the Friedmann 
magnitude-redshift relations in terms of $\Omega_{\rm{m}}$ and $\Omega_\Lambda$
separately, and not just in terms of their combination via $q_0$, for the 
redshift range of the P97 data one may adequately approximate the relation
by 
\begin{equation}
m_{\rm{B}} = {\cal M}_{\rm{B}} + 5 \log_{10}cz + 1.086 (1 - q_0) z  ,
\end{equation}
where
\begin{equation}
{\cal M}_{\rm{B}} = M_{\rm{B}} - 5 \log_{10} H_0 + 25  ,
\end{equation}
with the corresponding expression for the corrected B band
magnitudes. For reasons which will become clear when we consider the Stephani 
models, it is useful for us to write Eq. (14) in this form, in terms of
$q_0$. Since we will make use of similar expressions for the
Stephani universes, we construct for the Friedmann case the (reduced)
chi-squared statistic
\begin{equation}
\chi^2 = \frac{1}{n-1} \sum_{i=1}^{n} \left[
 \frac
{ m_{\rm{B}}^{\rm{obs}}(i) - m_{\rm{B}}^{\rm{pred}}(i) }
{\sigma(i)}
\right]^2  ,
\end{equation}
where $n$ is the number of SNIa,
$m_{\rm{B}}^{\rm{obs}}(i)$ and $\sigma(i)$ are respectively the observed
B band apparent magnitude and error estimate of the i$^{th}$ SNIa, and 
$m_{\rm{B}}^{\rm{pred}}(i)$ is the
predicted B band apparent magnitude of the i$^{th}$ SNIa, for a given value
of $q_0$, derived from equation 5 (or its equivalent for the corrected
magnitudes). Following P97 we adopt ${\cal M}_{\rm{B}} = -3.17 \pm 0.03$ and 
${\cal M}_{\rm{B,corr}} = -3.32 \pm 0.05$.

From Eqs. (14) and (16) it follows that ${\hat{q}}_0$,
the maximum likelihood (equivalently minimum $\chi^2$) estimate of $q_0$,
is given by:-
\begin{equation}
{\hat{q}}_0 = - \left[
\sum_{i=1}^{n}
\frac{x_i y_i} {\sigma^2(i)}
\right] \,
\left[
\sum_{i=1}^{n}
\frac{x_i^2} {\sigma^2(i)}
\right] ^{-1}  ,
\end{equation}
where
\begin{equation}
x_i = 1.086 z(i)
\end{equation}
and
\begin{equation}
y_i = m_{\rm{B}}^{\rm{obs}}(i) - {\cal M}_{\rm{B}}
- 5 \log cz(i) - 1.086 z(i)   .
\end{equation}
Substituting the appropriate values from P97 we find that
${\hat{q}}_0 = 0.48$ for the uncorrected magnitudes and
${\hat{q}}_0 = 0.50$ for the corrected magnitudes. Thus we see that
applying the magnitude corrections has negligible effect on the
best-fit value of $q_0$ for the P97 data.

Figure 1 shows the value of $\chi^2$, for $-0.5 < q_0 < 1$, both for the
uncorrected (solid line) and corrected data (dashed line). We can see from
this Figure that the corrected data consistently give a slightly better fit to
a Friedmann model than the uncorrected data for a wide range of values of
$q_0$. Both the uncorrected
and corrected data give an acceptable fit over the range $0 < q_0 < 1$ (and in
fact over a somewhat
wider range for the corrected data) but both give a poor
fit for large negative values of $q_0$.

Table 1 quantifies the goodness of fit of the SNIa data to a number of
Friedmann models with different values of the cosmological
parameters $\Omega_{\rm{m}}$, $\Omega_\Lambda$, and the age of the universe,
$t_0$ -- all for a Hubble constant $H_0 = 65$ km s$^{-1}$ Mpc$^{-1}$. Column
(3) shows the corresponding value of $q_0$, calculated from Eq. (1). Column
(5) gives the reduced $\chi^2$ of the fit to all seven SNIa, while
column (6) gives the reduced $\chi^2$ obtained using the five SNIa with 
corrected magnitudes.

\placetable{table1}

It is clear from Table 1 that one cannot obtain,
with $H_0 \simeq 65$ km s$^{-1}$ Mpc$^{-1}$,
an acceptable fit to either the corrected or uncorrected data
and at the same time ensure an age of the universe in excess of 14 Gyr. We 
discuss the situation for other values of $H_0$ below.

\subsection{Stephani Model II}

We now compare the SNIa data with the predicted magnitude-redshift relations of
the Stephani models. We consider first Model II, and the relation given by
Eq. (3). If we compare Eqs. (3) and (14) we see that in the limit as
$z \rightarrow 0$ these equations are identical if and only if
\begin{equation}
\tau_0 = \frac{2}{3} H_0^{-1}   .
\end{equation}
In other words for nearby SNIa, Stephani Model II predicts the same linear
redshift-magnitude relation as do Friedmann models, and with an age of the
universe equal to two thirds times the inverse of the Friedmann Hubble
constant. This is precisely the age of a Friedmann universe which is flat with
a zero cosmological constant. In particular, if $H_0 \sim 65$
km s$^{-1}$ Mpc$^{-1}$
then {\em independent of the value of the parameter, $\alpha$, } the age of the
universe $\tau_0$ in Model II is approximately 10 Gyr, which certainly 
appears to be
too low to be consistent with astrophysical age determinations. Hence it would
seem that Model II is not particularly useful in resolving the current age
conflict since the age is inextricably linked to the value of the
Friedmann Hubble constant: as soon as the latter is specified then so too is
the age of Model II.

The link between the magnitude-redshift relation for Model II and the
Friedmann case is, nonetheless, interesting for the following reason.
Note that Eq. (3) may be rewritten as
\begin{eqnarray}
m_{\rm{B}} & = &  M_{\rm{B}} + 25 - 5\log_{10} {\frac{2}{3\tau_{0}}} + 
5\log_{10}{cz}
\nonumber \\
      & + & 1.086 (1 - q_0) z  ,
\end{eqnarray}
where
\begin{equation}
q_0 =  \left( \frac{1}{2} - \frac{9}{8} c^2 \alpha
\tau_{0}^{\frac{4}{3}} \right)  ,
\end{equation}
which means that for any given age of the universe,
$\tau_0$, we can
choose the parameter, $\alpha$, so that the magnitude-redshift relation
for Model II is identical in form to Eq. (14), with
$\tau_0 = \frac{2}{3} H_0^{-1}$. The crucial difference is that, whereas in
the Friedmann case with $\Lambda = 0$, Eq. (20) implies that $q_0 = 0.5$,
in the Stephani case we still retain the freedom to specify
a relation which is equivalent to {\em any\/} value of $q_0$
by suitable
choice of $\alpha$. 

In particular, by choosing $\alpha < 0$ one can
obtain a magnitude-redshift relation which corresponds to a Friedmann model
with $q_0 > 0.5$. This is in full analogy to Friedmann models if the relation 
for the curvature of the Stephani models is taken into account 
(D\c{a}browski 1995, Eq. 2.14). It shows that 
the time dependent curvature index (Eq. 10)
for $\alpha < 0$ is positive (if cosmic time, $\tau > 0$), while for 
$\alpha > 0$ it is negative. The pressure (Eq. 12) is positive or negative for 
$\alpha$ being positive or negative respectively at the center of symmetry 
$r = 0$ and it diverges to either plus or minus 
infinity (depending on the values of other parameters) at the antipodal 
center of symmetry $ r \to \infty$ and the particles are either accelerated away 
or towards $r = 0$. 
Bearing in mind the effect of curvature (Eq. 10) of the models one can roughly 
say that the inclusion of non-uniform pressure mimics a flat Friedman model 
$q_0 = 1/2$ to become curved -- positively curved for $\alpha < 0$, and 
negatively curved for $\alpha > 0$.  

While in the Friedmann case $q_0 > 0.5$ would imply an
age of the universe $\tau_0 < \frac{2}{3} H_0^{-1}$, in the case of Model II
we still have $\tau_0 = \frac{2}{3} H_0^{-1}$.
Model II would, therefore, be of considerable interest if SNIa (or
other) observations were to suggest that $q_0 > 0.5$, which certainly 
cannot be ruled out on the
basis of the P97 data alone. As an illustrative (if somewhat extreme) example,
consider the case where $\Omega_{\rm{m}} = 2$ and $\Omega_{\Lambda} = 0$,
so that $q_0 = 1$ for the Friedmann model. As can be seen from Eq. (22)
the Friedmann  and Model II magnitude-redshift relations are identical when
\begin{equation}
\alpha = - \frac{4}{9} c^{-2} \tau_{0}^{-\frac{4}{3}}  .
\end{equation}
Whereas the age of the Friedmann model with $q_0 = 1$
would be reduced by about 15\% compared with the Einstein de Sitter age (i.e.
$\tau_0 \simeq 0.57 H_0^{-1}$), for the Stephani Model II we still have
$\tau_0 = \frac{2}{3} H_0^{-1}$. Although the scenario of $q_0 > 0.5$
appears highly unlikely, in view of a variety of other observations of large
scale structure and CMBR anisotropies, this serves as an interesting example
of how the Stephani models can be compatible with high redshift observations
over a larger region of parameter space than Friedmann models.

\subsection{Stephani Model I}

One of the reasons why Model II is not particularly useful as an extension of
the Friedmann case is that the effect of the non-uniform pressure (manifest
via the parameter $\alpha$) only becomes apparent at high redshift. The
situation with Stephani Model I is different, however. We can see from
Eq. (2) that the effect on the magnitude-redshift relation of the
non-uniform pressure parameter $a$ is immediate. In particular,
therefore, even at low $z$ Model I does not in general reduce trivially to
a specific Friedmann case.

Note that after setting the parameter $b$ from Model I equal to unity
(provided $d=0$) we can rewrite Eq. (2) to depend only upon the non-uniform 
pressure parameter $a$. Thus
\begin{eqnarray}
m_{\rm{B}} & = & M_{\rm{B}} + 25 
+ 5\log_{10}{ \left[cz \left(\frac{a\tau_{0}^2 + 
\tau_{0}}{2 a\tau_{0} + 1}\right)\right]} \nonumber \\
& + & 1.086 \left[ 1 + 4 \frac{\left( a \tau_{0}^2 + \tau_{0} 
\right)}{\left(2 a \tau_{0} + 1 \right)^2} \right] z  .
\end{eqnarray}

As a means of estimating what
range of values of $a$ and $\tau_0$ will give an
acceptable fit to the P97 data it is useful to note further that we may
recast Eq. (24) in the form
\begin{eqnarray}
m_{\rm{B}} & = & M_{\rm{B}} + 25 + 5 \log_{10} cz - 5 \log_{10} \tilde{H}_0
\nonumber \\
           & + & 1.086 (1 - \tilde{q}_0) z  ,
\end{eqnarray}
where
\begin{equation}
\tilde{H}_0 = \frac{2a\tau_{0} + 1}{a\tau_{0}^2 + \tau_{0}}
\end{equation}
and
\begin{equation}
\tilde{q}_0 = - 4a \frac{a\tau_{0}^2 + \tau_{0}}{(2 a\tau_0 + 1)^2}  .
\end{equation}

Eq. (24) now takes the same functional form as Eq. (14),
as was similarly pointed out in
D\c{a}browski (1995), with $\tilde{H}_0$ and $\tilde{q}_0$ replacing
$H_0$ and $q_0$. We can think of $\tilde{H}_0$ (which is one third of the
expansion scalar $\Theta$ of the model) and $\tilde{q}_0$ as a
generalised Hubble parameter and deceleration parameter
which are related to the
age of the universe in a different way from the Friedmann case. The key
question of interest here is therefore whether one can construct
generalised parameters, $\tilde{H}_0$ and $\tilde{q}_0$, which are in good
agreement with the P97 data but which correspond to a value of $\tau_0$
which exceeds that Friedmann age with $H_0 = \tilde{H}_0$ and 
$q_0 = \tilde{q}_0$.
The fact that we can write the Model I redshift-magnitude relation in the form
of Eq. (25) confirms, however, that our choices of $\tau_0$ and $a$
are certainly {\em not\/} arbitrary. Combinations of $\tau_0$ and $a$
which give a large negative value of $\tilde{q}_0$, for example, would clearly
be incompatible with the SNIa Hubble diagram -- just as was the case for
Friedmann models with $q_0 << 0$.

In order to estimate the parameters $\tau_0$ and $a$ we
construct the reduced chi-squared statistic:-
\begin{equation}
\chi^2 = \frac{1}{n-2} \sum_{i=1}^{n} \left[
 \frac
{ m_{\rm{B}}^{\rm{obs}}(i) - m_{\rm{B}}^{\rm{pred}}(i;\tau_0, a) }
{\sigma(i)}
\right]^2  ,
\end{equation}
where $n$ is the number of SNIa and $m_{\rm{B}}^{\rm{pred}}(i;\tau_0,a)$
is obtained for the i$^{th}$ SNIa from Eqs. (25), (26) and (27). We
determine $M_{\rm{B}}$ from Hamuy et al. (1996), adopting their
best-fit value of $H_0 = 63.1$ km s$^{-1}$ Mpc$^{-1}$ determined from four
local calibrating SNIa. Thus,
\begin{equation}
M_{\rm{B}} \equiv {\cal M}_{\rm{B}} + 5 \log H_0 - 25 = -19.17 \pm 0.03
\end{equation}
and
\begin{equation}
M_{\rm{B,corr}} \equiv
{\cal M}_{\rm{B,corr}} + 5 \log H_0 - 25 = -19.32 \pm 0.05  .
\end{equation}

The dependence of Eq. (28) on $\tau_0$ and $a$ is non-linear,
making a plot of the surface $z = \chi^2(\tau_0,a)$
difficult to interpret. We therefore consider slices through this surface.
Moreover for plots of $\chi^2$ at constant $\tau_0$ it is useful to plot
$\chi^2$ as a function of $a^{-1}$.
Figure 2 shows $\chi^2$ as a function of $a^{-1}$ for $\tau_0 = 13$ Gyr,
using the uncorrected P97 data. The behaviour of
$\chi^2$ is seen to be rapidly varying for values of $a^{-1}$ around
zero, but is essentially flat for all $|a^{-1}| \gte 0.3$. Thus, provided
$|a| \lte $ 3 km$^2$ s$^{-2}$ Mpc$^{-1}$, we see that the goodness of fit
of Model I to the data is essentially independent of the value of $a$, and
depends only on $\tau_0$. A very similar curve is obtained for
the corrected magnitudes.

We can understand the rapidly varying behaviour of $\chi^2$ for small values of
$|a^{-1}|$ by considering the behaviour
of $\tilde{H}_0$ and $\tilde{q}_0$ in Eqs. (26) and (27). We see that
when $a^{-1} = - 2 \tau_0$ we have $\tilde{H}_0 = 0$ and $\vert \tilde{q}_0
\vert \rightarrow \infty$
\footnote{This is a special situation, and moreover one which
contradicts the observations since it would 
mean that the present day is exactly the turning point of the
cosmic evolution. Following Eqs. (7) and (8)  
the energy density at the moment $\tau_0 = -1/2a$ is $8\pi G/c^4 \varrho = 48a^2$ 
and the pressure $8\pi G p = \varrho (-1/3 + 1/6 r^2)$. In the 
Friedman limit $a \rightarrow 0$ we have $\tau_0 \rightarrow \infty$, 
and $p = \varrho = 0$ which is an everlasting, empty, flat and static 
universe -- a subcase of little or no physical interest.}, so that 
$\chi^2 \rightarrow \infty$.
It therefore follows that for $\tau_0 = 13$ Gyr $ = 13.26 \times 10^{-3}$
s Mpc km$^{-1}$ a singular value of $\chi^2$ occurs when
$a^{-1} \simeq -0.026$, and $\chi^2$ varies very rapidly close to this value.
However, the range of very small $a^{-1}$ is not of interest to us
since it deviates too far from Friedmann models.

Figures 3a and 3b show plots of $\chi^2$ as a function of $a^{-1}$ but
now with $\tau_0 = 15$ Gyr, for the corrected and uncorrected magnitudes 
respectively. A narrower range of values of $a^{-1}$ is shown, in order to
better illustrate the behaviour of $\chi^2$ for small $|a^{-1}|$. We find
that $\chi^2$ again tends to infinity when $a^{-1} = - 2 \tau_0$, and is
again essentially flat for all $|a^{-1}| \gte 0.3$. Note also that the
asymptotic value of $\chi^2$ is a little smaller than that for $\tau_0 = 13$
Gyr, and moreover that there exists a narrow range of values of $a$ for which
$\chi^2$ dips appreciably below its asymptotic value.

Figures 4a and 4b, on the other hand,
show plots of $\chi^2$ as a function of $\tau_0$ for 
$a = - 1.0$, for the uncorrected and corrected magnitudes respectively.
This value of $a$ is chosen to be representative of the asymptotic
behaviour of $\chi^2$; essentially the same plots would be obtained for
all $|a| \lte 3$ km$^2$ s$^{-2}$ Mpc$^{-1}$. We can see that Model I gives 
a good fit to the data for $\tau_0$ in the range 13 to 15 Gyr.

Figures 5a and 5b show the values of
$\tilde{H}_0$ and $\tilde{q}_0$ respectively as a function of $\tau_0$,
again for the representative value of $a = - 1.0$.
Also shown for comparison are the best-fit values
of $H_0 = 63.1$ km s$^{-1}$ Mpc$^{-1}$
and $q_0 = 0.5$, obtained from Hamuy et al. (1996) and section 4.1
respectively. We see from Figure 5b that $\tilde{q}_0$ is almost independent
of $\tau_0$ over the range shown, increasing from $q_0 \simeq 0.04$
($\tau_0$ = 10 Gyr) to $q_0 \simeq 0.08$ ($\tau_0$ = 20 Gyr).
The dependence of $\tilde{H}_0$ on $\tau_0$ is rather more pronounced,
however: good agreement with the Hamuy et al. value is found in the age
range 14 to 16 Gyr.

The behaviour of $\tilde{H}_0$ and $\tilde{q}_0$ in Figures 5a and 5b
makes sense
when we consider the form of Eqs. (26) and (27) for $a\tau_0 << 1$.
To first order in $a\tau_0$ these reduce to
\begin{equation}
\tilde{H}_0 = \frac{1}{\tau_0} \left( 1 + a\tau_0 \right)
\end{equation}
and
\begin{equation}
\tilde{q}_0 = - 4 a \tau_{0}  .
\end{equation}
Thus we see that as $a\tau_0 \rightarrow 0$,
$\tilde{H}_0 \rightarrow \tau_0^{-1}$ and
$\tilde{q}_0 \rightarrow 0$.

The potential usefulness of Model I is now apparent. In the limit where
$a \tau_0 \rightarrow 0$, the age of the universe in this model
is increased by 50\% compared with the Einstein de Sitter age, giving
for example $\tau_0 = 15$ Gyr (compared with only 10 Gyr) for $H_0 \sim 65$.   
Of course when $a \tau_0 \rightarrow 0$, $\tilde{q}_0 \rightarrow 0$ also,
so that the more meaningful comparison is that between Model I and a
Friedmann model with $q_0 = 0$. Taking $\Omega_{\rm{m}} = 0.3$, $q_0 = 0$
and $H_0 = 65$ gives a Friedmann age of only 12.5 Gyr, however, so that the
Model I age is still 2.5 Gyr greater than the Friedmann age. Taking larger
values for the matter density results in a bigger difference between the
Friedmann and Model I ages.

Figures 5a and 5b are indicative of the limiting behaviour of
$\tilde{H}_0$ and $\tilde{q}_0$ for small $a \tau_0$. As we already
remarked for the $\chi^2$ plots, one obtains similar plots for all 
$|a| \lte 3$ km$^2$ s$^{-2}$ Mpc$^{-1}$, with $\tilde{H}_0$
strongly varying as a function of $\tau_0$ but $\tilde{q}_0$ much more
weakly dependent on $\tau_0$. The smaller the value of $|a|$ the
closer $\tilde{q}_0$ lies to zero at given $\tau_0$ -- as is obvious from
Eq. (32). The {\em sign\/} of $a$ does have some bearing on the goodness of 
fit of the model to the P97 data, however. Although it can be seen from Figure 1
and Table 1 that the current data give an acceptable fit for $q_0 = 0$,
the fit rapidly deteriorates for negative values of $q_0$. Thus, if $a > 0$
(i.e. a high pressure region at $r = 0$ away from which particles are
accelerated) then $\tilde{q}_0 \rightarrow 0$ from below, and a value of
$a = 3$ km$^2$ s$^{-2}$ Mpc$^{-1}$ would imply $\tilde{q}_0 \simeq -0.2$ for
$\tau_0 = 15$ Gyr, which gives only a marginally acceptable
fit to the P97 data.
If $a < 0$, on the other hand (i.e. a low pressure region at $r = 0$ towards
which particles are accelerated) then one can obtain much better fits: e.g.
$a = -3$ km$^2$ s$^{-2}$ Mpc$^{-1}$ implies $\tilde{q}_0 \simeq 0.2$ for
$\tau_0 = 15$ Gyr.

We have emphasised the limiting behaviour of Model I, for small $a$, 
in order to make clear that the usefulness of the model is a fairly robust 
result and is not too sensitive to the exact value of $a$ which is
chosen -- although it is true that negative values of $a$ are favoured. 
It is particularly noteworthy that the limit $\vert a \vert \lte 3$, may be 
considered as the restriction on this parameter from the observational data 
and allows us not to be too far from the range where Friedmann models are valid.
In other words one can
obtain a good fit to the P97 data, with a significantly larger age, but 
without requiring that Model I departs too much from a Friedmann model.

According to Eq. (9), $|a| \lte 3$ km$^2$ s$^{-2}$ Mpc${-1}$ translates
to a limit on the value of acceleration scalar of
$\vert \dot{u} \vert \lte 0.66 \times 10^{-10} \, r$ Mpc$^{-1}$, where $r$ is 
the radial coordinate in the model. 
Notice that, although the pressure is different from zero at the center $r = 0$ 
the effect of acceleration is not detectable at the center since the 
acceleration scalar (and of course vector) vanishes there. 

If we allow slightly larger negative values for the non-uniform pressure
parameter, $a$, then by careful choice of $a$ and $\tau_0$ we can
obtain fits which give significant positive values of $\tilde{q}_0$, while
retaining a significant difference between the Stephani and Friedmann ages.
Although these fits require slightly more `fine tuning' they are
clearly in much closer agreement with the best-fit Friedmann value 
of $q_0 = 0.5$. Some examples of fits of this type are given in Table 2. 
The final 
two columns of Table 2  show the age, $\tau_{\rm{F}}$, of the universe in a 
Friedmann model with $\Lambda = 0$ and with 
$\Omega_{\rm{m}} + \Omega_{\Lambda} = 1$ respectively.

\placetable{table2}
 
Some general trends are evident from Table 2. Note that in all cases 
we see that as $\tau_0$ 
increases and $|a|$ becomes smaller, the values of $\tilde{H}_0$
and $\tilde{q}_0$ are both reduced and the goodness of fit to the corrected
and uncorrected data gradually deteriorates. 
For $\tau_0 \geq 16$ Gyr, the goodness of fit quickly becomes unacceptably 
large: although by suitable choice of
$a$ one can ensure that $\tilde{q}_0$ remains positive, the value of
$\tilde{H}_0$ also reduces and overall the fit deteriorates. Further decrease
in $|a|$, for $\tau_0 \geq 16$ Gyr, increases the value of $\tilde{H}_0$, but
pushes $\tilde{q}_0$ closer to zero, so that the goodness of fit remains poor.
This behaviour can also be easily seen from Eqs. (31) and (32).

It would seem, therefore, that an age of $\tau_0 = 15 - 16$ Gyr represents 
the upper
age limit from Model I with the P97 data -- at least if one adopts the
SNIa calibration of $H_0$. Moreover, if subsequent analysis of larger samples 
of SNIa serve to tighten the limits on a positive value of $\tilde{q}_0$, 
then this limiting age could perhaps be reduced to $\tau_0 \sim 14$ Gyr.
The important point to note, however, is that in this case the age limits on
Friedmann models would {\em also\/} be reduced. As can be seen from Table 2,
a value of $\tilde{q}_0 \sim 0.5$ can be well fitted by Model I
with $\tau_0 = 14$ Gyr, which still represents an increase in the age of
the universe of more than 3 Gyr compared with the $q_0 = 0.5$
Friedmann model with either zero cosmological constant or critical density.

\section{Conclusions}

In this paper we have considered the two exact non-uniform pressure
spherically symmetric Stephani universes, and have compared the
redshift-magnitude relations derived for these models in D\c{a}browski (1995)
with the recent SNIa observations of P97. We have investigated the extent
to which, by suitable choice of the Stephani model parameters, we may obtain
good fits of the P97 data to the predicted redshift-magnitude relations
but for universes which are older than their Friedmann counterparts. 

We emphasize that we have considered only the case of centrally placed 
observers, which results in having zero pressure gradient in our location. 
Although this is clearly a special case, it is mathematically the simplest 
possibility and merits consideration first. It can be extended relatively 
simply to the case of a non-centrally placed observer using the formulae given
in Section 5 of D\c{a}browski (1995). However, such a generalisation introduces 
additional model parameters which make apparent magnitude a function of both 
redshift and direction the sky. In principle we could have estimated these
parameters in this paper, but the small size of the P97 sample would make such a 
parameter fit statistically meaningless. Indeed, even for the case of a centrally
placed observer, ideally one should consider a much larger supernovae sample. We 
will consider the more general case in future, when the number of observed 
supernovae has significantly increased.

We have found that the age of the universe in Stephani Model II is, in
fact, independent of the non-uniform pressure 
parameter, $\alpha$, and is equal to
the age of an Einstein de Sitter Friedmann model, i.e.
$\tau_0 = \frac{2}{3} H_0^{-1}$. This model would be of considerable interest
if the total density of the universe were greater than the critical density,
since the age of the corresponding Friedmann model would then be {\em less\/}
than the Einstein de Sitter age. Since there exists no compelling
observational evidence to suggest that the universe is closed, however,
Model II is of limited use as it would in general predict an age which was
smaller than its Friedmann counterpart with the same value of the Hubble
constant.

We have shown that Stephani Model I would be of considerably greater
interest, however. We have found that the redshift-magnitude relation
predicted for Model I can be expressed in terms of two parameters: the
age, $\tau_0$, of the universe and the 
non-uniform pressure parameter $a$. One can write the redshift-magnitude
relation in exactly the same form as in the Friedmann case, introducing
an effective Hubble parameter, $\tilde{H}_0$, and deceleration parameter,
$\tilde{q}_0$, which are non-linear functions of $a$ and $\tau_0$.
We have shown that for a wide range of different values of $a$ we
can obtain good fits to the P97 data for a universe of age up to 15 Gyr,
which is typically two or three Gyr greater than the corresponding
Friedmann model.
These fits are quite robust, requiring only that 
$|a| \lte 3$ km$^2$ s$^{-2}$ Mpc$^{-1}$, which gives the value of the 
acceleration scalar $\dot{u}$ of the order $\vert \dot{u} \vert \lte
0.66 \times 10^{-10} \, r $ Mpc
$^{-1}$, where $r$ is the radial coordinate of the model. Then, although the 
pressure is not zero at the center of symmetry $r = 0$, the effect of 
acceleration is non-detectable at the center since the acceleration scalar 
vanishes there. However, this effect is easily extracted from the 
redshift-magnitude relation since neighbouring galaxies are not situated at the 
center and they necessarily experience acceleration.

The above robust fits are for the limiting case where the product
$a \tau_0$ is small, and imply an effective deceleration parameter, $\tilde{q}_0$,
close to zero -- a value which is certainly not as yet ruled out by the P97
data, although the fit to a value of $\tilde{q}_0 = 0.5$ is
currently somewhat better.
By some fine-tuning of the Model I parameters, one can obtain good fits
with $\tilde{q}_0 \sim 0.5$ and $\tau_0 \sim 14$ Gyr. While an age of
only 14 Gyr may still be in conflict with astrophysical age
determinations, the conflict is considerably worse for Friedmann models:
the age of an $H_0 \sim 65$, $q_0 \sim 0.5$ critical density Friedmann
universe is only $\sim 10$ Gyr, and for closed models with $q_0 \sim 0.5$
the age is even smaller.

Thus, we find that Model I can give an age of the universe
which is consistenly and robustly between two and three
Gyr older than the oldest acceptable open or flat Friedmann models.

Since the preliminary results of P97 were first presented, there have been
several important developments in the measurement of fundamental cosmological
parameters. The recalibration of the RR Lyrae distance scale has revised age
estimates of the oldest globular clusters to $t_0 = 11.7 \pm 1.4$ Gyr (c.f.
Chaboyer et al. 1997). This undoubtedly lessens the conflict with the
standard ($\Omega_0 = 1$, $\Lambda = 0$) cosmological model --
particularly if one argues for a value of $H_0 \sim 55$ (c.f. Tammann 1996).
If one requires a `gestation period' of around 1 Gyr between the Big Bang
and the formation of the first globular clusters, however, then agreement with
the standard model is still only marginal -- even for $H_0 = 55$ -- and
open Friedmann models would appear to be favoured.
Since we have argued in this paper that the P97 data does not yet exclude
models with $q_0 \sim 0$, it is only fair to point out that open Friedmann
models with $\Lambda \neq 0$, $q_0 \sim 0$ and $H_0 \sim 55$ offer a
comfortable resolution of the age problem, in the light of the Chaboyer et al.
results. Adopting, for example, $\Omega_{\rm{m}} = 0.5$, 
$\Omega_{\Lambda} = 0.25$ and $H_0 = 55$ gives a Friedmann age of 
$\tau_{\rm{F}} = 14.0$ Gyr.

It is important to recognise, however, that this agreement rests crucially
upon the value of $H_0$. If, instead, one adopts the most recent estimate
of $H_0$ from the HST distance scale Key Project: $H_0 = 73 \pm 6 \pm 8$
km s$^{-1}$ Mpc$^{-1}$ (Freedman 1996), the above Friedmann age reduces to 
only $10.6$ Gyr,
and for the standard model (with $q_0 = 0.5$) is only 8.9 Gyr. Moreover, the 
impact on $H_0$ of 
the HIPPARCOS recalibration of the LMC distance modulus has recently been 
shown by Madore and Freedman (1997) to be less significant than had 
previously been reported (c.f. Feast \& Catchpole 1997). Thus it would seem
that rumours of the end of the age problem are perhaps somewhat premature.
If, indeed, the value of $H_0$ lies close to that obtained by the HST Key 
Project then we note that the Stephani models considered here can still give
an age of up to approximately 12.5 Gyr, with $q_0 \sim 0.5$, and 13.4 Gyr,
with $q_0 \sim 0$. While the data clearly do not yet present a case for
abandoning Friedmann models, equally they do not rule out the 
possible need to do so in the future -- when bounds on $H_0$ and $q_0$ are 
tightened -- and the Stephani models investigated here could indeed prove
to be very important.

In this paper we have considered only a particular class of inhomogeneous
models, in order to illustrate their potential usefulness in addressing the
apparent conflict between the observed values of the Friedmann model
parameters. In future work we will extend our treatment to a wider class of
non-Friedmann models and test their compatibility with the Hubble diagram
of high-redshift objects and other cosmological observations. Such a
comparison will be greatly enhanced by having larger samples of distant
SNIa -- a development which the modern observing strategies adopted by P97
and other groups will shortly provide.

As for the further progress in our comparison of the Stephani models with 
astronomical data one should of course investigate
such standard tests as galaxy number 
counts, the angular size-redshift relation or microwave background 
anisotropies as far as to adopt the second order terms in redshift $z^2$. 
This can indeed be done after some relatively tedious calculations, again 
using the powerful power series methods originally given by 
Kristian \& Sachs (1966, see also Ellis 1971 for more detailed discussion). 
Many of these issues have been studied previously for Friedmann models 
(e.g. D\c{a}browski \& Stelmach 1986, 1987; Stelmach, Byrka \&  
D\c{a}browski 1990) and will be the subject of future work.   

\acknowledgements
The authors thank John Barrow, Richard Barrett and Chris Clarkson for many 
useful discussions. MAH acknowledges the receipt of a PPARC personal research 
fellowship and a PPARC research assistantship. MPD acknowledges the support of 
the Royal Society and the hospitality of the Astronomy Centre at the
University of Sussex.

\clearpage
\begin{figure}
%
%
\figcaption[]{The value of $\chi^2$, for $-0.5 < q_0 < 1$, 
both for the uncorrected (solid line) and corrected data (dashed line).}
\end{figure}

\begin{figure}
%
%
\figcaption[]{Plot of the variation of $\chi^2$ with $a^{-1}$, 
for a fixed value of $\tau_0 = 12$ Gyr, obtained from a comparison of the apparent magnitudes predicted by Stephani Model I with the uncorrected magnitudes 
of P97.}
\end{figure}

\begin{figure}
%
%
\figcaption[]{Plot of the variation of $\chi^2$ with $a^{-1}$, 
for a fixed value of
$\tau_0 = 15$ Gyr, obtained from a comparison of the apparent magnitudes
predicted by Stephani Model I with the magnitudes of P97. Figure 3a is for
the uncorrected magnitudes and Figure 3b is for the corrected magnitudes.}
\end{figure}

\begin{figure}
%
%
\figcaption[]{Plot of the variation of $\chi^2$ with $\tau_0$, for a fixed 
value of $a = -1.0$, obtained from a comparison of the apparent magnitudes
predicted by Stephani Model I with the magnitudes of P97. Figure 4a is for
the uncorrected magnitudes and Figure 4b is for the corrected magnitudes.}
\end{figure}

\begin{figure}
%
%
\figcaption[]{Plot of the effective Friedmann parameters,
$\tilde{H}_0$ and $\tilde{q}_0$, as a function of $\tau_0$,
for a fixed value of
$a = -1.0$. Dotted lines indicate the best-fit Friedmann values for $H_0$
(Figure 5a) from Hamuy et al. (1996) and $q_0$ (Figure 5b) from section
4.1.}
\end{figure}

\clearpage
\begin{deluxetable}{ccrccc}
\footnotesize
\tablecaption{Goodness of fit, expressed as a reduced $\chi^2$,
of the SNIa data to Friedmann models with different
combinations of $\Omega_{\rm{m}}$ and $\Omega_{\Lambda}$. Column (3) gives the
corresponding value of $q_0$ and column (4) the age, $t_0$, of the universe
assuming $H_0 = 65$ km s$^{-1}$ Mpc$^{-1}$. Columns (5) and (6) respectively give
the reduced $\chi^2$ of the fit to all seven SNIa and to the five SNIa with
corrected magnitudes. \label{table1}}
\tablewidth{0pt}
\tablehead{
\colhead{$\Omega_{\rm{m}}$} & \colhead{$\Omega_{\Lambda}$} & 
\colhead{$q_0$}             & \colhead{$t_0$} & \colhead{$\chi^2$} &
\colhead{$\chi^2_{\rm{corr}}$} \nl
 & & & Gyr & &
}
\startdata
1.0 & 0.0  &  0.5  & 10.0 & 0.87 & 0.37 \nl
0.5 & 0.0  &  0.25 & 11.3 & 0.99 & 0.55 \nl
0.3 & 0.0  &  0.15 & 12.2 & 1.12 & 0.70 \nl
0.2 & 0.0  &  0.1  & 12.7 & 1.20 & 0.79 \nl
0.3 & 0.15 &  0.0  & 12.9 & 1.40 & 1.02 \nl
0.2 & 0.1  &  0.0  & 13.4 & 1.40 & 1.02 \nl
0.3 & 0.25 & -0.1  & 12.8 & 1.68 & 1.32 \nl
0.5 & 0.5  & -0.25 & 12.5 & 2.09 & 1.77 \nl
0.3 & 0.7  & -0.55 & 14.5 & 3.31 & 3.06 \nl
0.2 & 0.8  & -0.7  & 16.2 & 4.07 & 3.86 \nl
\enddata
\end{deluxetable}

\clearpage
\begin{deluxetable}{cccccccc}
\footnotesize
\tablecaption{Fits of Stephani Model I universes with significantly
positive values of $\tilde{q}_0$ to the P97 data. The final two
columns show the age, $\tau_{\rm{F}}$,
of the universe in a Friedmann model with the same values of
$H_0$ and $q_0$, with $\Lambda = 0$
and with $\Omega_{\rm{m}} + \Omega_{\Lambda} = 1$ respectively. \label{table2}}
\tablewidth{0pt}
\tablehead{
\colhead{$\tau_0$} & \colhead{$a$}              & \colhead{$\tilde{H}_0$} & 
\colhead{$\tilde{q}_0$} & \colhead{$\chi^2$} & 
\colhead{$\chi^2_{\rm{corr}}$} & \colhead{$\tau_{\rm{F}}$ $(\Lambda=0)$} &
\colhead{$\tau_{\rm{F}}$ $(\Omega_{\rm{m}} + \Omega_{\Lambda} = 1)$} \nl
 Gyr               & km$^2$ s$^{-2}$ Mpc$^{-1}$ & km s$^{-1}$ Mpc$^{-1}$  &
                        &                    &
                               & Gyr                                      & 
 Gyr
}
\startdata
13.00 & -10.00 & 63.7 & 0.86 & 1.67 & 1.07 &  9.1 &  9.5 \nl
13.25 &  -8.33 & 64.4 & 0.67 & 1.40 & 0.79 &  9.5 &  9.8 \nl
13.50 &  -7.14 & 64.5 & 0.55 & 1.24 & 0.64 &  9.9 & 10.0 \nl
13.75 &  -6.67 & 63.8 & 0.51 & 1.15 & 0.56 & 10.2 & 10.2 \nl
14.00 &  -6.25 & 63.0 & 0.48 & 1.12 & 0.58 & 10.4 & 10.3 \nl
14.25 &  -5.55 & 62.6 & 0.42 & 1.14 & 0.65 & 10.8 & 10.6 \nl
14.50 &  -5.00 & 62.0 & 0.38 & 1.22 & 0.77 & 11.1 & 10.8 \nl
14.75 &  -4.55 & 61.4 & 0.34 & 1.32 & 0.94 & 11.4 & 11.0 \nl
15.00 &  -4.17 & 61.3 & 0.27 & 1.48 & 1.14 & 11.9 & 11.2 \nl
15.25 &  -3.85 & 60.0 & 0.29 & 1.66 & 1.42 & 12.0 & 11.4 \nl
15.50 &  -3.33 & 59.6 & 0.25 & 1.88 & 1.72 & 12.4 & 11.6 \nl
15.75 &  -3.12 & 58.8 & 0.23 & 2.14 & 2.07 & 12.7 & 11.8 \nl
16.00 &  -0.35 & 58.1 & 0.22 & 2.43 & 2.45 & 12.9 & 12.0 \nl
\enddata
\end{deluxetable}
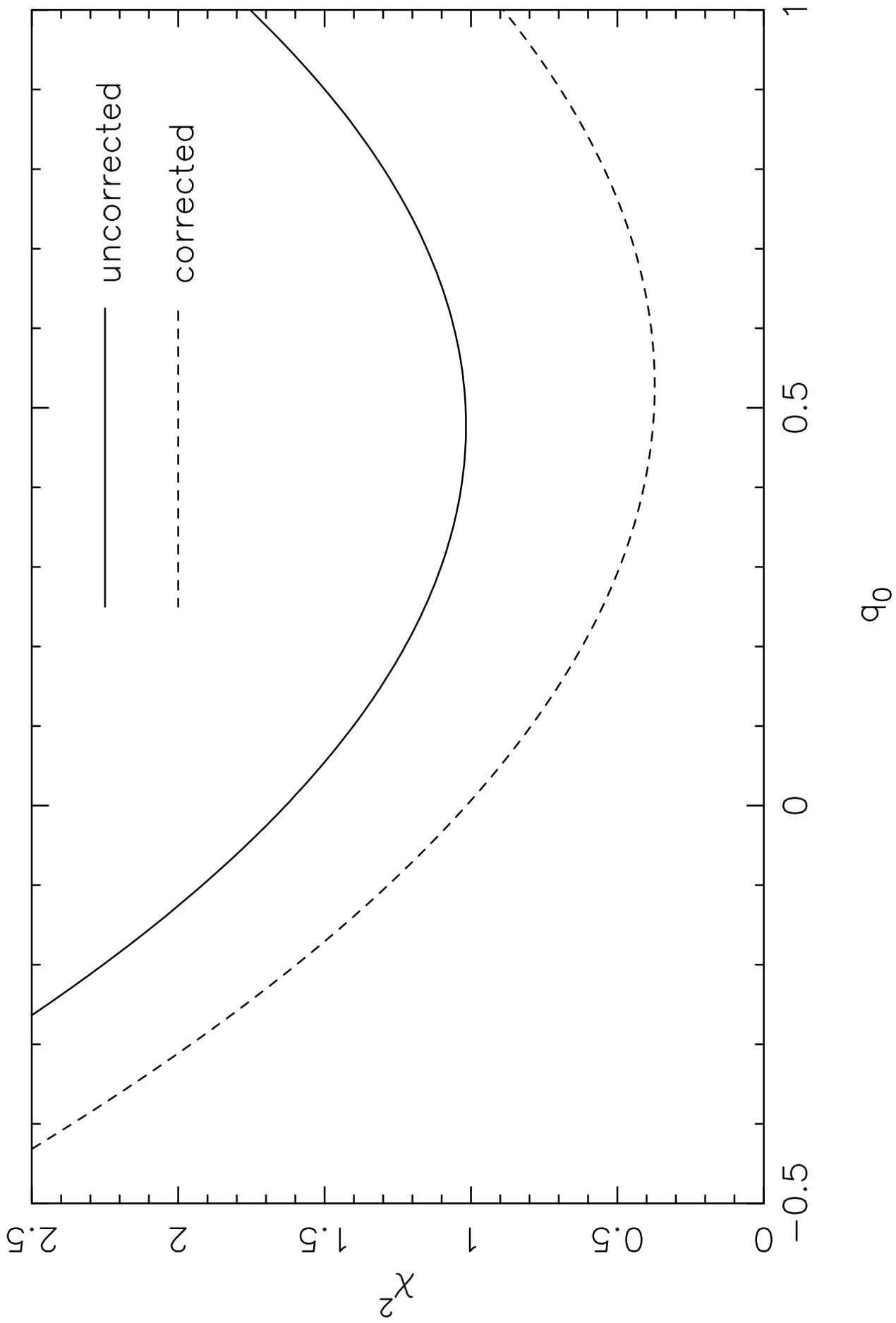
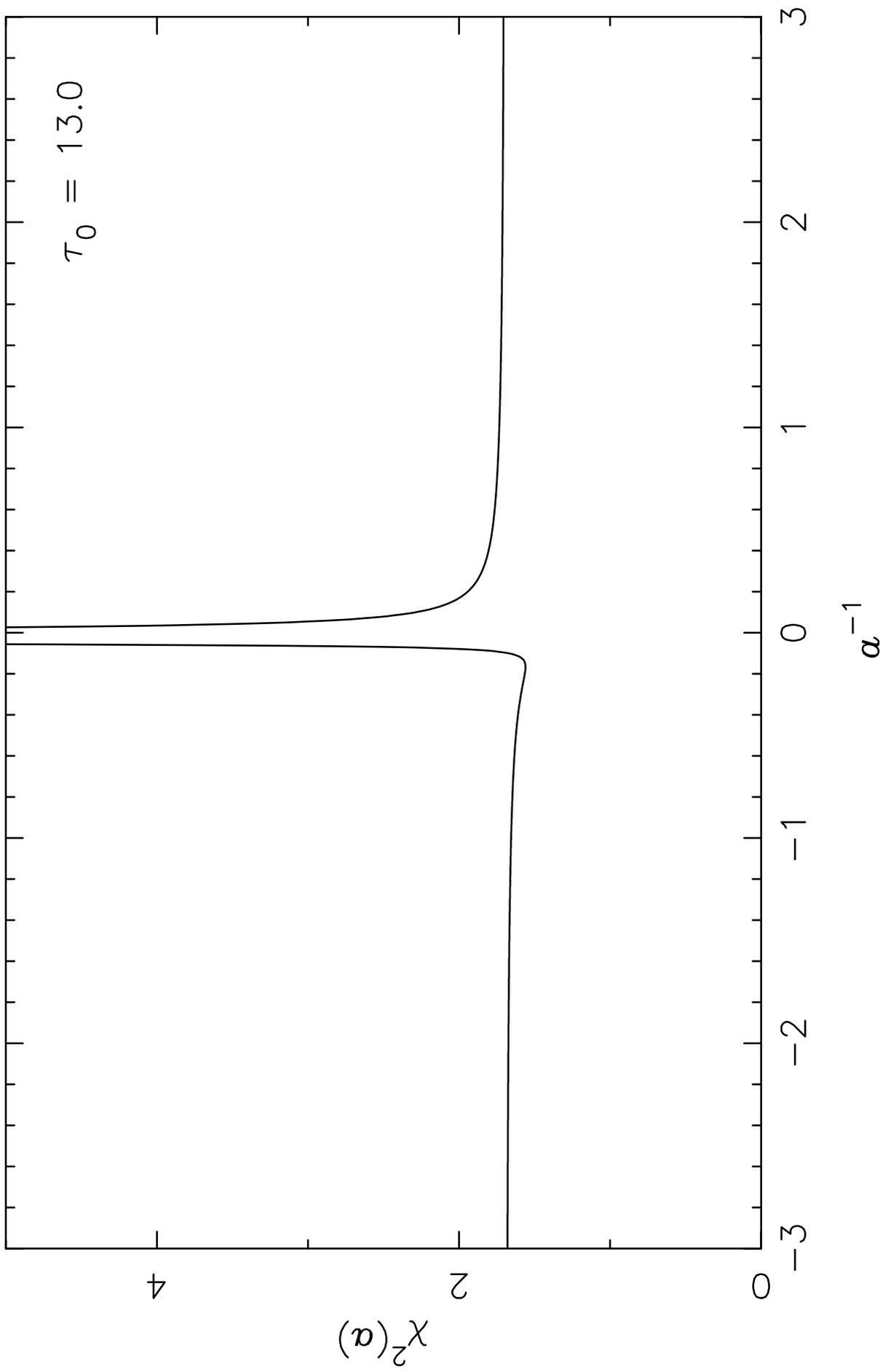
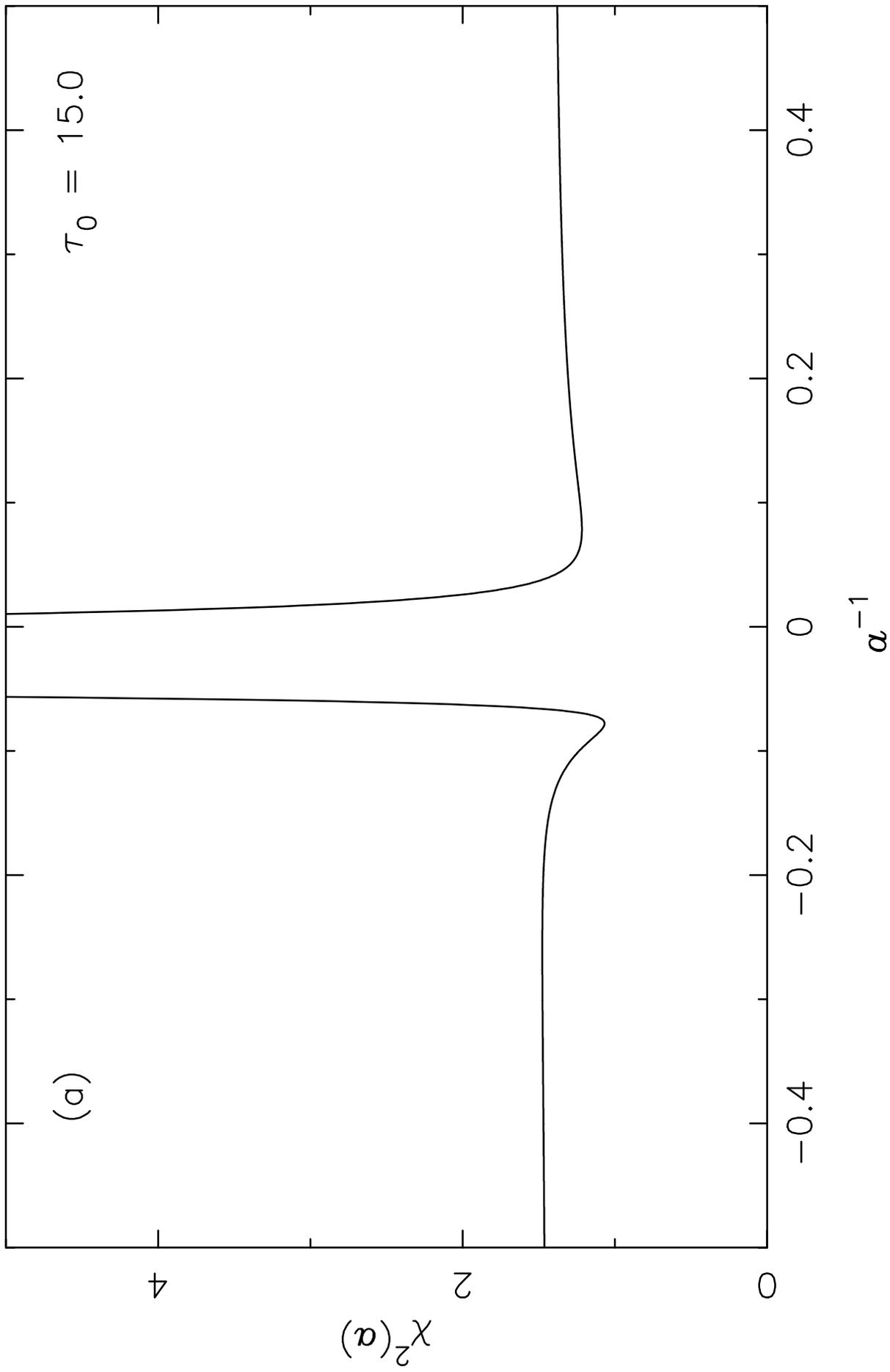
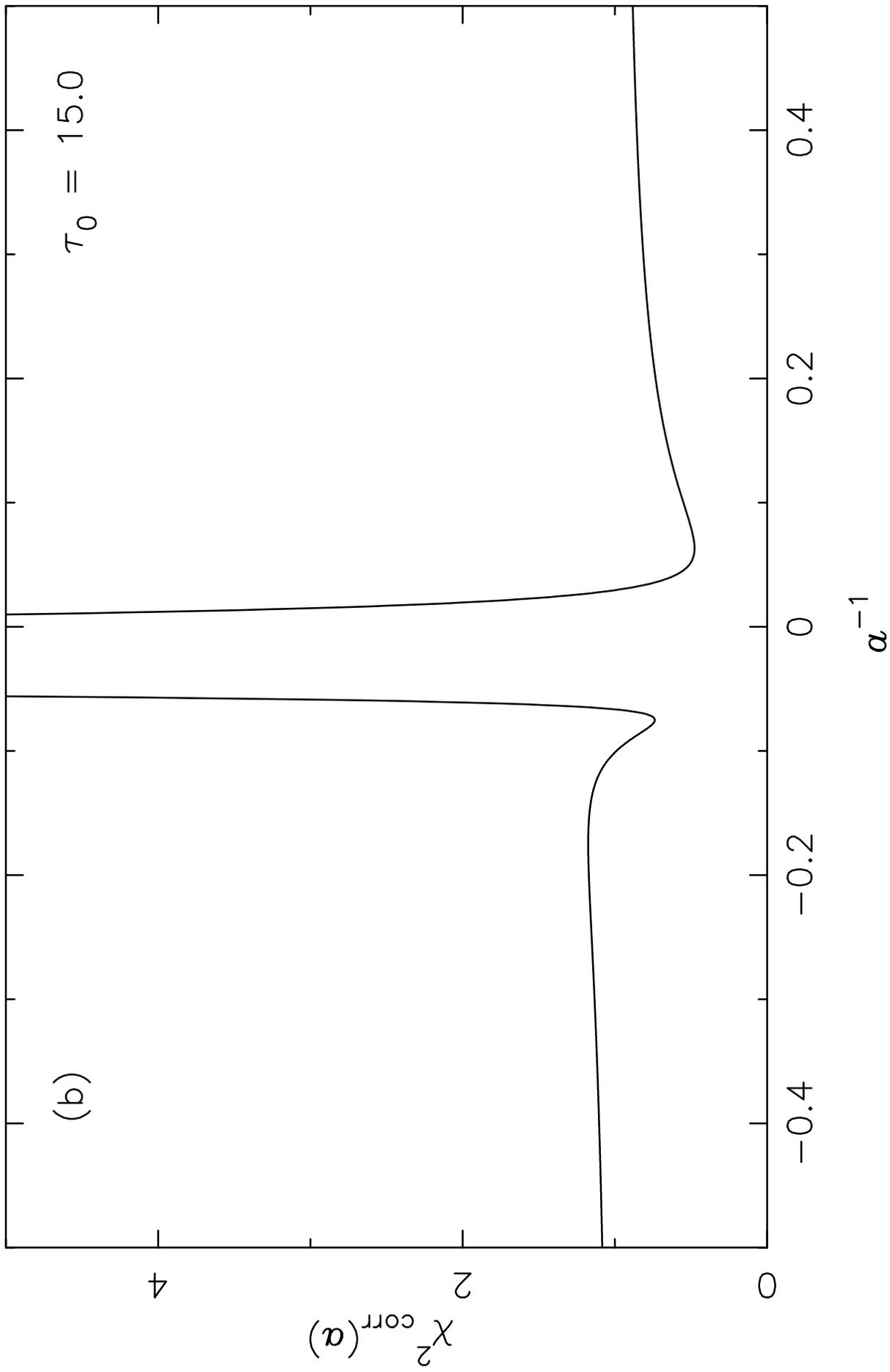
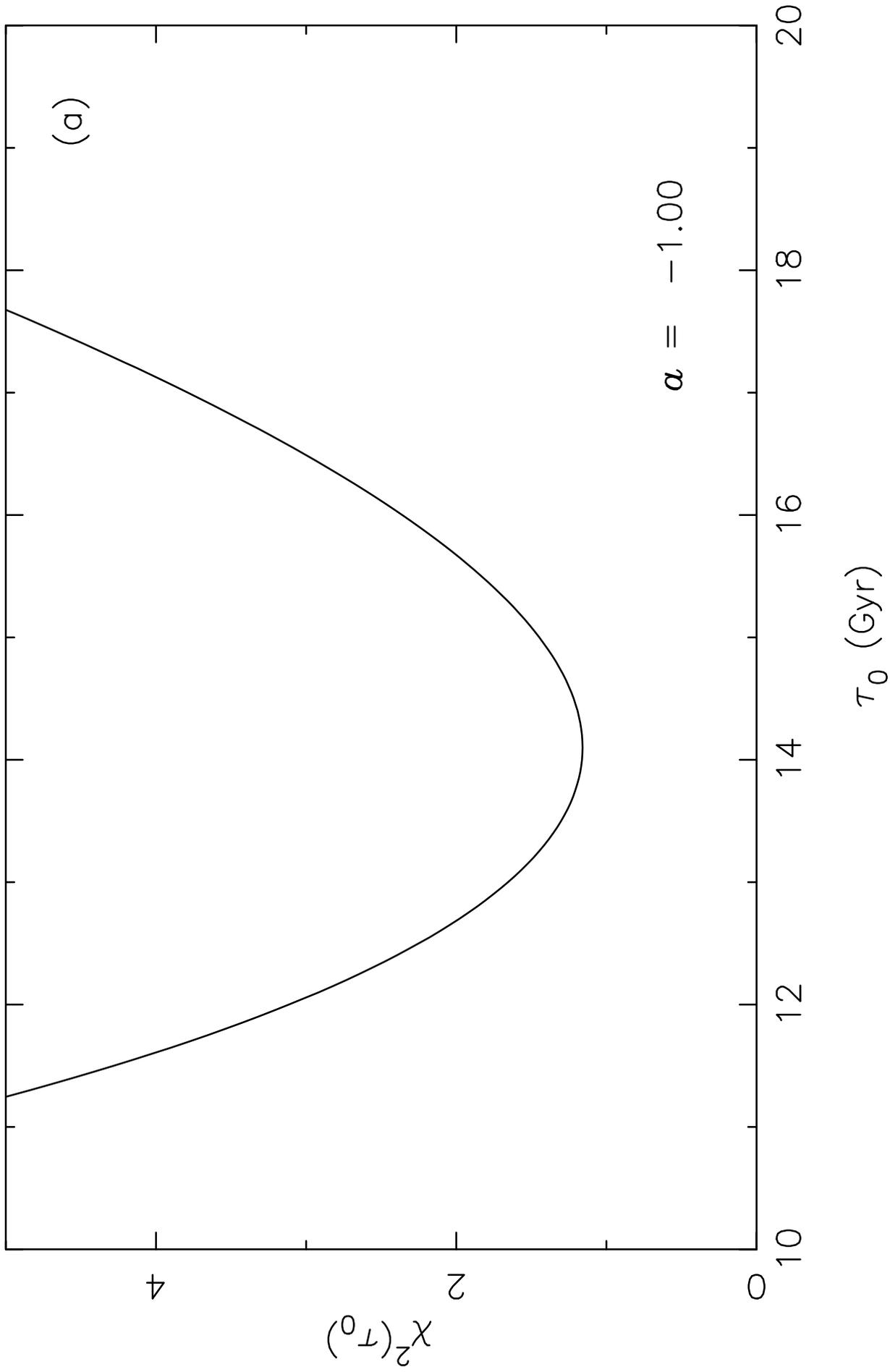
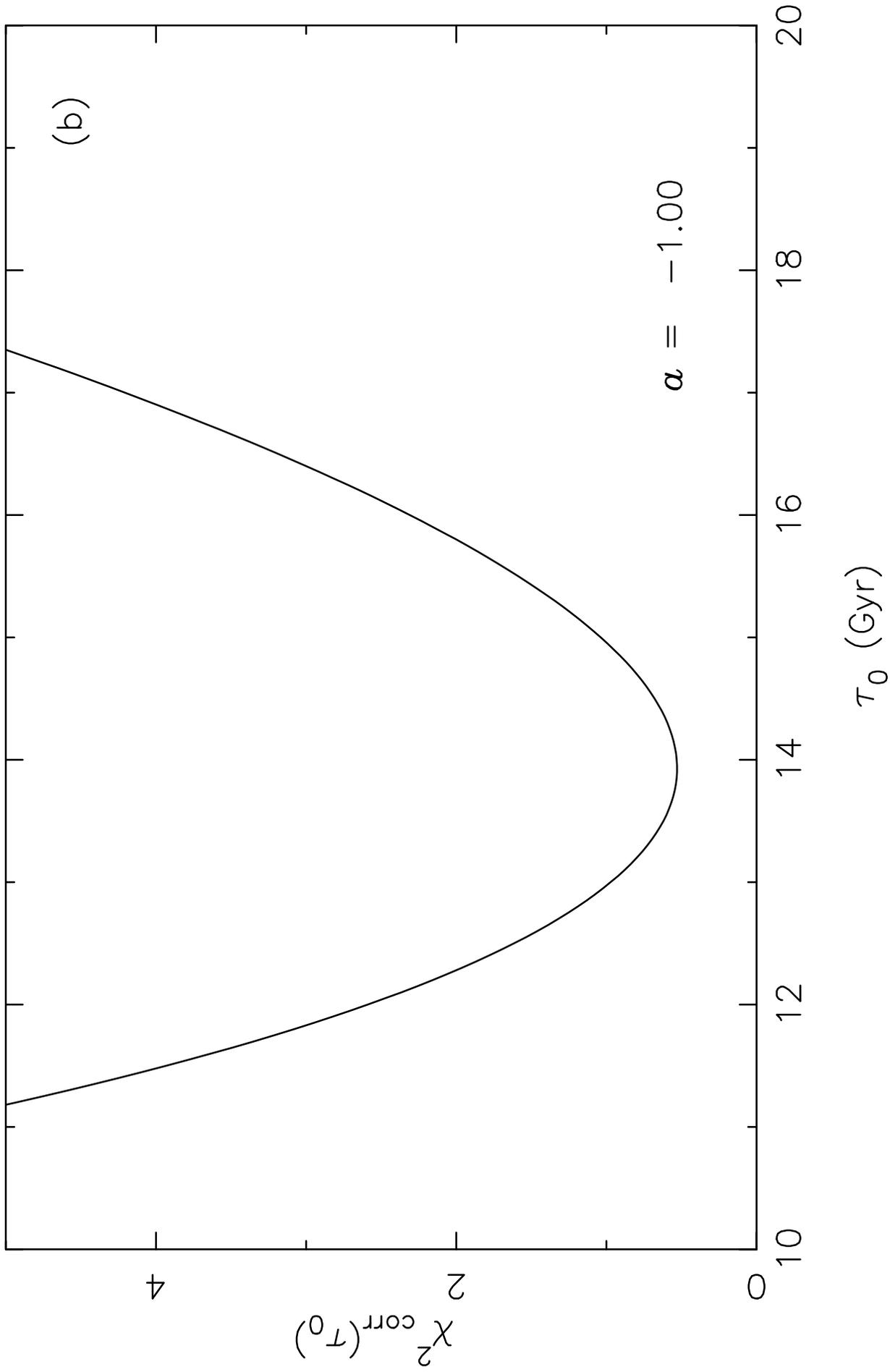
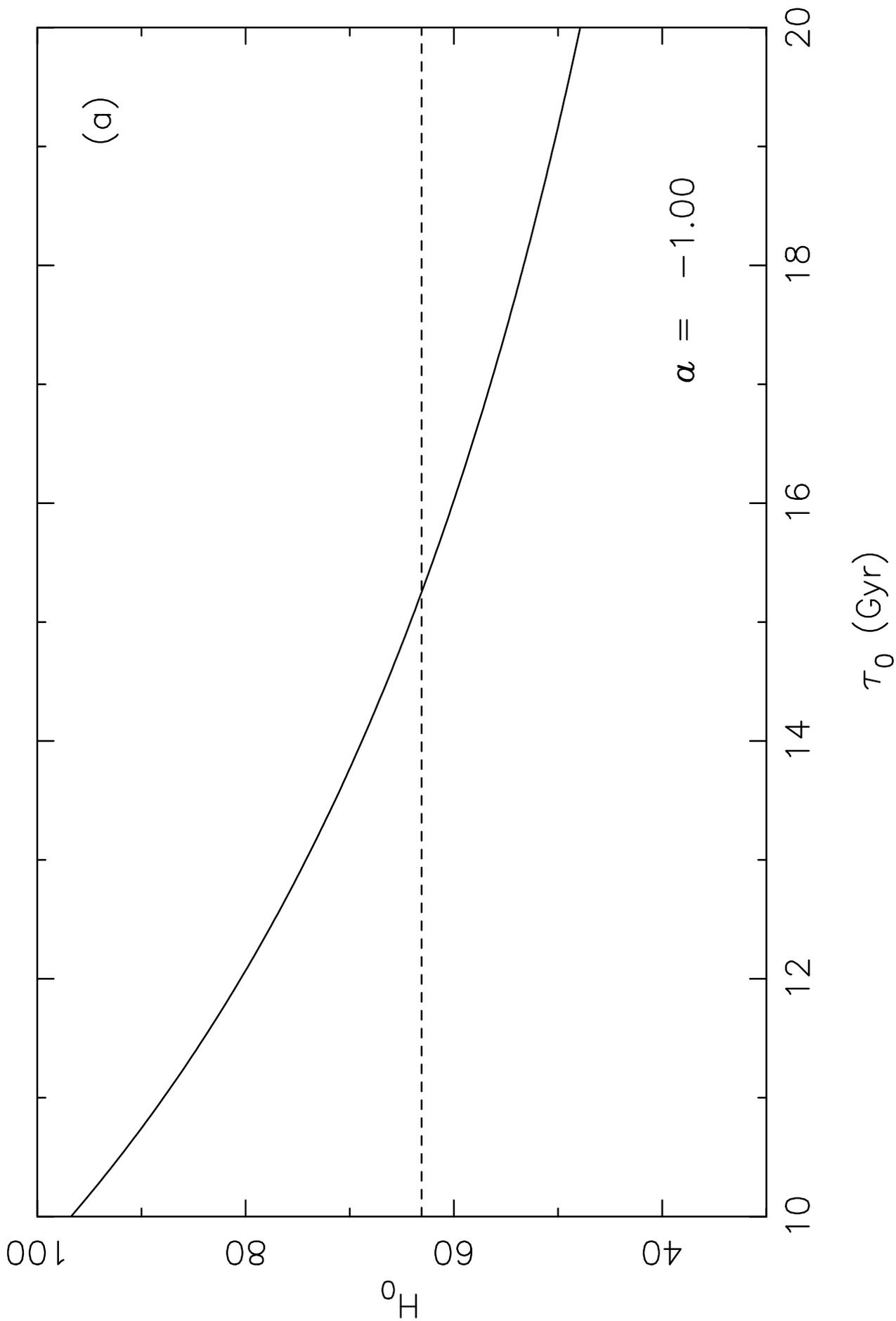
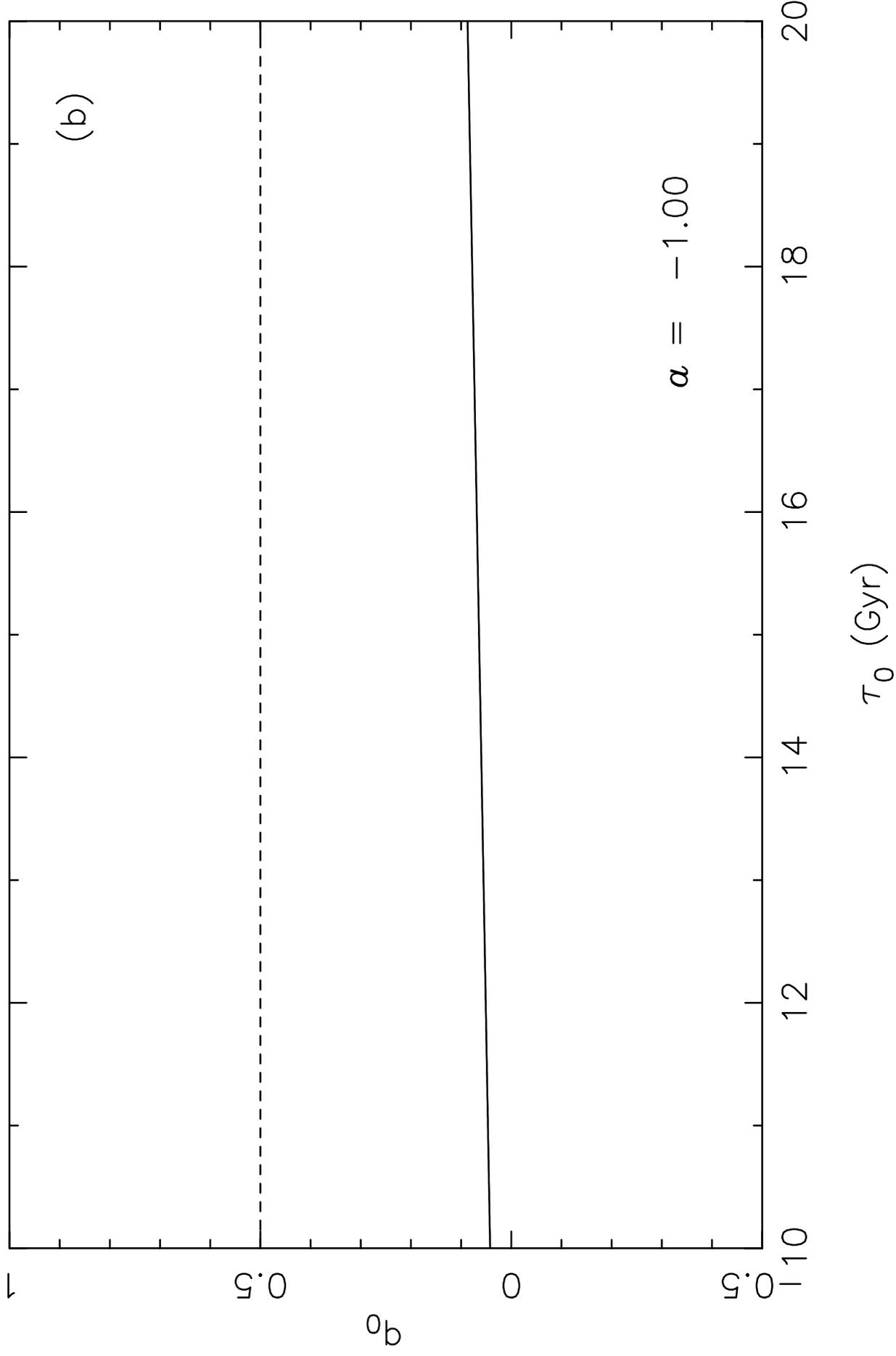
\end{document}